\documentclass[prd,twocolumn,showpacs,floatfix,superscriptaddress,nofootinbib]{revtex4}
\usepackage{graphicx}
\usepackage{amssymb}
\usepackage{dcolumn}
\usepackage{amsmath}
\usepackage{bm}
\usepackage{textcomp}
\usepackage{epstopdf}
\usepackage{color}
\usepackage{ulem}
\usepackage[toc,page,title,titletoc,header]{appendix}
\usepackage[colorlinks,
            linkcolor=blue,
            anchorcolor=blue,
            citecolor=blue
            ]{hyperref}
\usepackage{txfonts}
\usepackage{caption}
\usepackage{multirow}

\begin{document}
\title{Correlated fluctuations near the QCD critical point}

\begin{abstract}
In this paper, we introduce a freeze-out scheme for the dynamical models near the QCD critical point through coupling
the decoupled classical particles with the order parameter field. With a modified distribution
function that satisfies specific static fluctuations, we calculate the correlated fluctuations of net protons
on the hydrodynamic freeze-out surface. A comparison with recent STAR
data shows that our model calculations could roughly reproduce energy
dependent cumulant $C_4$ and $\kappa \sigma^2$ of net protons through tuning the related parameters. However, the calculated
 $C_2$ and $C_3$ with both Poisson and Binomial baselines
are always above the experimental data due to the positive contributions
from the static critical fluctuations. In order to qualitatively and quantitatively describe all the related experimental data,
the dynamical critical fluctuations and more realistic non-critical fluctuation baselines should be investigated
in the near future.
\end{abstract}

\author{Lijia Jiang}
\affiliation{Department of Physics and State Key Laboratory of Nuclear Physics and
Technology, Peking University, Beijing 100871, China}
\author{Pengfei Li}
\affiliation{Department of Physics and State Key Laboratory of Nuclear Physics and
Technology, Peking University, Beijing 100871, China}
\author{Huichao Song}
\email{Huichao Song: huichaosong@pku.edu.cn}
\affiliation{Department of Physics and State Key Laboratory of Nuclear Physics and
Technology, Peking University, Beijing 100871, China}
\affiliation{Collaborative Innovation Center of Quantum Matter, Beijing 100871, China}
\affiliation{Center for High Energy Physics, Peking University, Beijing 100871, China}
\maketitle

\section{Introduction}.

The Beam Energy Scan (BES) program at the Relativistic Heavy Ion Collisions (RHIC)
aims to explore the QCD phase diagram and study the QCD phase transition~\cite{Aggarwal:2010cw}. At small
chemical potential region, lattice simulations demonstrate that the QCD
phase transition is a rapid but smooth crossover~\cite{Karsch:2003jg,Aoki:2006we,Aoki:2009sc,Bazavov:2014pvz}.
At large chemical potential and medium
temperature region, different effective models show that the QCD phase
transition is the first order with a clear phase boundary~\cite{Klevansky:1992qe,Fukushima:2003fw,Roberts:1994dr,Berges:2000ew}.
Correspondingly, a critical point is located at the end of the first order
phase boundary, leading to unique critical phenomena of the QCD matter
~\cite{Stephanov:2004wx}.

In the vicinity of the critical point, the correlation length $\xi$ and
the magnitude of fluctuations become divergent in a static and infinite medium.
Detailed calculations have shown that
the correlated fluctuations between two particles are proportional to the
square of the correlation length: $\left\langle \left( \delta N\right)
^{2}\right\rangle _{c}\sim \xi ^{2}$~\cite{Stephanov:1998dy,Stephanov:1999zu}, higher cumulants of the correlated fluctuations are more
sensitive to the correlation length with $\left\langle \left( \delta
N\right) ^{3}\right\rangle _{c}\sim \xi ^{4.5}$ and $\left\langle \left( \delta
N\right) ^{4}\right\rangle _{c}\sim \xi ^{7}$~\cite{Stephanov:2008qz,Athanasiou:2010kw}.
It was also found that the fluctuations of net protons are related to the baryon susceptibilities~\cite{Hatta:2003wn,Kitazawa:2012at}
- the quantities that are generally used to evaluate the critical fluctuations in Lattice QCD and effective field theories.
Therefore, the higher cumulants for the net-proton distributions are treated as the
main observables to search the critical point in experiment.

Recently, the STAR collaboration has measured the energy dependent moments of net-proton
multiplicity distributions in Au+Au collisions at $\sqrt{s_{NN}}$ = 7.7, 11.5, 19.6,
27, 39, 62.4 and 200 GeV~\cite{Aggarwal:2010wy,Adamczyk:2013dal,Luo:2015ewa}. With the maximum transverse momentum increased from
$0.8$ to $2$ GeV, the cumulant ratio $\kappa \sigma ^{2}$ for net protons shows large
deviation from the Poisson expectations and presents an obvious non-monotonic behavior in
central Au+Au collisions~\cite{Luo:2015ewa}, which shows the
potential to discover the QCD critical point.

To quantitatively study these experimental measurements, one needs to
develop dynamical models near the QCD critical point. Based on
linear sigma model, Paech and her collaborators have established the framework of
chiral hydrodynamics through coupling the hydrodynamical
equations for quarks with the evolution equations for the long wavelength
mode of the sigma field\cite{Paech:2003fe}. Later on, the Frankfurt group further
developed chiral hydrodynamics with dissipation and noise~\cite{Nahrgang:2011mg,Nahrgang:2011vn}, and with the contributions
from the Polyakov loop~\cite{Herold:2013bi}.
Besides the dynamical evolution, an equation of state with a critical point has also been constructed
and been applied to pure hydrodynamic simulations~\cite{Nonaka:2004pg,Asakawa:2008ti}.

As far as we know, most of the dynamical models near the QCD critical point only
focus on the dynamical evolution of the bulk matter~\cite{Paech:2003fe,Nahrgang:2011vn,Nahrgang:2011mg,Herold:2013bi}. Without a proper
treatment of the freeze-out procedure, these models can not be used to directly calculate the fluctuations
of produced hadrons as measured in experiment.  In this paper, we will introduce
a freeze-out scheme for the dynamical models near the QCD critical point using a modified distribution function.
To avoid the numerical complexities, we will not investigate the dynamical critical phenomena from
a realistic dynamical evolution, but concentrate on studying the static critical fluctuations of emitted classical particles
from a pure hydrodynamic freeze-out surface. We will demonstrate that, with Poisson and Binomial
statistical fluctuation baselines, the static critical fluctuations from our model calculations
can roughly reproduce energy dependent cumulant $C_4$ and $\kappa \sigma^2$ of net protons as measured in experiment.
However, the calculated $C_2$ and $C_3$ with both Poisson and Binomial baselines
are always above the experimental data due to the positive contributions
from the static critical fluctuations. In order to qualitatively and quantitatively describe all the related
experimental data, the dynamical critical fluctuations and more realistic non-critical fluctuation baselines should be investigated
in the near future.
\\[0.15in]

\section{The formalism}.

In this section,  we will introduce a freeze-out scheme for the evolution
of the bulk matter near the critical point, and then deduce the formulism to calculate
the 2nd, 3rd and 4th cumulants for the correlated fluctuations of the classical particles emitted from the
hydrodynamic freeze-out surface near $T_c$ with the presence of an external order parameter field. We emphasis that our formalism
aims to investigate the static critical phenomena near the QCD critical point,
which does not involve the dynamical evolution of the order parameter field. We also neglect
the hadronic scatterings and resonance decays below $T_c$, but leave the investigations of such effects
to the future study.

In traditional hydrodynamics, the classical particles emitted from the freezeout
surface can be calculated through the Cooper-Frye formula~\cite{PhysRevD.10.186,Kolb:2000sd}:
\begin{equation}
E\frac{dN}{d^{3}p}=\int_{\Sigma }\frac{p_{\mu }d\sigma ^{\mu }}{2\pi ^{3}}%
f\left( x,p\right) ,
\end{equation}%
where $\Sigma $ is the freezeout hyper-surface, $d\sigma ^{\mu }$ is the
normal vector on the freezeout surface, and $f\left(x,p\right)$ is the distribution
function for the classical particles. For Boltzmann statistics, its  co-variant
form is written as $f\left(x,p\right) =e^{-(p^{\mu }u_{\mu }-\mu )/T}$  where $p^{\mu }$ is the
four-momentum of the classical particles (with $p_0=\sqrt{m_0^2+\mathbf{p}^2}$, $m_0$ is
the physics mass of the  particles), $u^{\mu }$ is the four-velocity of the fluid
and $T$ is the decoupling temperature on the freeze-out surface.

In the vicinity of the critical point, we assume the classical particles still
satisfy some modified distributions $f(x,p)$, but with variable
effective masses that fluctuate in position space
through interacting with the correlated fluctuating order parameter
field $\sigma$. For example, the proton interacts with the sigma field through
the $\sigma NN$ coupling,
its effective mass can be written into two parts: $m=m_{0}+\delta m$. $m_{0}$ is
the physical mass, $\delta m=g\sigma (x)$ is the variable mass induced by the
sigma filed $\sigma (x)$~\footnote{We have shifted the average value of the sigma field to zero.
$\sigma (x)$ here represents the fluctuations of the sigma field.}which strongly fluctuates
near the critical point~\cite{Stephanov:2011pb}.
$g$ is the coupling between the proton and the sigma field. In this
way, the correlated fluctuations of the sigma field are translated to the
distribution function $f(x,p)$ of the classical particles~\footnote{Based on similar idea,
Ref.~\cite{Stephanov:2011pb} has derived the standard critical fluctuations in
a static and infinite medium~\cite{Stephanov:2008qz} with a fluctuating momentum-space
distribution function $f(p)$.}.

With the modified distribution function and the variable effective mass,
one can, in principle, calculate the critical fluctuations of produced particles through the Cooper-Frye formula
in a dynamical model near the critical point, e.g., chiral hydrodynamics.
However, such calculations involve event-by-event simulations of the coupled evolution equations for
the sigma field and for the bulk matter. To compare with the experimental data, especially,
the higher cumulants of net protons, one needs to generate the fluctuating
sigma fields $\sigma (x)$ that satisfies specific 2-point, 3-point and 4-point correlators. Unfortranately,
this has not been numerically realized in current dynamical simulations.

In this article, we will perform event-averaged calculations for the correlated fluctuations of particles emitted
from the hydrodynamic freeze-out surface through expanding the modified equilibrium distribution functions
~\footnote{Such equilibrium distribution functions is a stationary solution for the Boltzmann equation
with the presence of an external sigma field~\cite{Stephanov:2009ra}.\\[0.15in]}
to the linear order of $\sigma (x)$~\cite{Stephanov:2011pb, Stephanov:2009ra}. With such expansion, the distribution
function is written as:
\begin{equation}
f=f_{0}+\delta f=f_{0}\left( 1-g\sigma /\left( \gamma T\right) \right) ,
\end{equation}
where $f_{0}$ is the traditional equilibrium distribution function: $f_{0}=f\left(
m_{0}\right) $, $\delta f$ represents the fluctuation, and $\gamma =\frac{p^{\mu }u_{\mu }}{m}$ is the covariant Lorentz factor.
 With such expansion, the 2-point, 3-point and 4-point correlators of $\delta f$ are expressed as:
\begin{eqnarray}
\left\langle \delta f_{1}\delta f_{2}\right\rangle _{c} &=&\frac{f_{01}f_{02}%
}{\gamma _{1}\gamma _{2}}\frac{g^{2}}{T^{2}}\left\langle \sigma _{1}\sigma
_{2}\right\rangle _{c}, \\
\left\langle \delta f_{1}\delta f_{2}\delta f_{3}\right\rangle _{c} &=&-%
\frac{f_{01}f_{02}f_{03}}{\gamma _{1}\gamma _{2}\gamma _{3}}\frac{g^{3}}{%
T^{3}}\left\langle \sigma _{1}\sigma _{2}\sigma _{3}\right\rangle _{c}, \\
\left\langle \delta f_{1}\delta f_{2}\delta f_{3}\delta f_{4}\right\rangle
_{c} &=&\frac{f_{01}f_{02}f_{03}f_{04}}{\gamma _{1}\gamma _{2}\gamma
_{3}\gamma _{4}}\frac{g^{4}}{T^{4}}\left\langle \sigma _{1}\sigma _{2}\sigma
_{3}\sigma _{4}\right\rangle _{c}.
\end{eqnarray}%
note that $\left\langle {}\right\rangle _{c}$ denotes the expectation values
of the connected correlators since disconnected correlators do not
contribute to the related cumulants of final produced hadrons.


The above correlators Eqs.(3-5) require detailed expressions for the correlations of the sigma field.
To simplify the calculations, we neglect the feedback interactions from particles and only consider the static
critical fluctuations of the sigma field. Following Ref.~\cite{Stephanov:2008qz, Stephanov:2011pb},
we use the probability distribution $P[\sigma ]$ with a potential involve
cubic and quartic terms to calculate the correlators of the
sigma field, which is written as:
\begin{equation}
P[\sigma ]\sim \exp \left\{ -\Omega \left[ \sigma \right] /T\right\} ,
\end{equation}
with
\begin{equation}
\Omega \left[ \sigma \right] =\int d^{3}x\left[ \frac{1}{2}\left( \nabla
\sigma \right) ^{2}+\frac{1}{2}m_{\sigma }^{2}\sigma ^{2}+\frac{\lambda _{3}%
}{3}\sigma ^{3}+\frac{\lambda _{4}}{4}\sigma ^{4}\right] ,
\end{equation}%
With this probability distribution, the correlators of the
sigma field at the tree diagram level are:%
\begin{eqnarray}
\left\langle \sigma _{1}\sigma _{2}\right\rangle _{c} &=&TD\left(
x_{1}-x_{2}\right) , \\
\left\langle \sigma _{1}\sigma _{2}\sigma _{3}\right\rangle _{c}
&=&-2T^{2}\lambda _{3}\int d^{3}zD\left( x_{1}-z\right) D\left(
x_{2}-z\right)   \notag \\
&&\times D\left( x_{3}-z\right) ,
\end{eqnarray}%
\begin{eqnarray}
\left\langle \sigma _{1}\sigma _{2}\sigma _{3}\sigma _{4}\right\rangle _{c}
&=&-6T^{3}\lambda _{4}\int d^{3}zD\left( x_{1}-z\right) D\left(
x_{2}-z\right)   \notag \\
&&\times D\left( x_{3}-z\right) D\left( x_{4}-z\right)   \notag \\
&&+12T^{3}\lambda _{3}^{2}\int d^{3}u\int d^{3}vD\left( x_{1}-u\right)
D\left( x_{2}-u\right)   \notag \\
&&\times D\left( x_{3}-v\right) D\left( x_{4}-v\right) D\left( u-v\right) .
\end{eqnarray}%
where the propagator $D\left( x-y\right) =\frac{1}{4\pi r}e^{-m_\sigma r}$, $r=|x-y|$, $m_\sigma$ is
the mass of the sigma field, and the corresponding correlation length is $\xi =1/m_\sigma$.
\\[-0.05 in]

With the above detailed expressions, one can calculate the cumulants of
produced hadrons through integrating Eqs.(3-5) on
the hydrodynamic freeze-out surface:
\begin{align}
\left\langle \left( \delta N\right) ^{2}\right\rangle _{c}=& \left( \frac{%
1}{\left( 2\pi \right) ^{3}}\right) ^{2}\prod_{i=1,2}\left( \int \frac{1%
}{E_{i}}d^{3}p_{i}\int_{\Sigma _{i}}p_{i\mu }d\sigma _{i}^{\mu }d\eta
_{i}\right)  \notag \\
& \times \frac{f_{01}f_{02}}{\gamma _{1}\gamma _{2}}\frac{g^{2}}{T^{2}}%
\left\langle \sigma _{1}\sigma _{2}\right\rangle _{c}, \\
\left\langle \left( \delta N\right) ^{3}\right\rangle _{c}=& \left( \frac{%
1}{\left( 2\pi \right) ^{3}}\right) ^{3}\prod\limits_{i=1,2,3}\left(
\int \frac{1}{E_{i}}d^{3}p_{i}\int_{\Sigma _{i}}p_{i\mu }d\sigma _{i}^{\mu
}d\eta _{i}\right)  \notag \\
& \times \frac{f_{01}f_{02}f_{03}}{\gamma _{1}\gamma _{2}\gamma _{3}}\left(
-1\right) \frac{g^{3}}{T^{3}}\left\langle \sigma _{1}\sigma _{2}\sigma
_{3}\right\rangle _{c}, \\
\left\langle \left( \delta N\right) ^{4}\right\rangle _{c}=& \left( \frac{%
1}{\left( 2\pi \right) ^{3}}\right) ^{4}\prod\limits_{i=1,2,3,4}\left(
\int \frac{1}{E_{i}}d^{3}p_{i}\int_{\Sigma _{i}}p_{i\mu }d\sigma _{i}^{\mu
}d\eta _{i}\right)  \notag \\
& \times \frac{f_{01}f_{02}f_{03}f_{04}}{\gamma _{1}\gamma _{2}\gamma
_{3}\gamma _{4}}\frac{g^{4}}{T^{4}}\left\langle \sigma _{1}\sigma _{2}\sigma
_{3}\sigma _{4}\right\rangle _{c}.
\end{align}
In the following calculations, we denote these cumulants
of critical fluctuations $\left\langle \left( \delta N\right) ^{2}\right\rangle _{c}$,
$\left\langle \left( \delta N\right) ^{3}\right\rangle _{c}$
and $\left\langle \left( \delta N\right) ^{4}\right\rangle _{c}$
as $C_2^{critical}$, $C_3^{critical}$ and $C_4^{critical}$, respectively.

In Eqs.(11-13), the effects from the dynamical evolution of the bulk matter
have been partially accounted through
these multi-dimensional integrations on the hydrodynamic freeze-out surface.
However, our formulism still belongs to
the category of static critical fluctuations since the input correlators, Eqs.(8-10),
are static correlators without involving a time evolution of the sigma field.  If replacing the related integrations
of Eqs.(11-13) by the integrations over the whole position space,
the standard formula for a static and infinite medium deduced by Stephanov in 2009~\cite{Stephanov:2008qz}
can be reproduced (please see the appendix for details). Recently, Mukherjee and his collaborators~\cite{Mukherjee:2015swa}
have found that the cumulants of the sigma field could change their sign during the
dynamical evolution. However, their method can not be directly
implemented to our formulism since the related calculations only consider the zero mode of sigma field,
which already erase the needed spacial information.

\section{Set ups}.

Eqs.(11-13) require a (3+1)-dimensional freeze-out surface for the related multi-dimentional integrations.
To obtain such freeze-out surface, we implement the viscous hydrodynamic code VISH2+1~\cite{Song:2007fn} and extend its (2+1)-d
freeze-out surface to the longitudinal direction within the measured
momentum rapidity range $|y|<0.5$ through the rapidity
correlations between momentum and space~\cite{Song:2010aq}. The hydrodynamic simulations start at $\tau
_{0}=0.6$ $\mathrm{fm/c}$ with smooth initial conditions generated from the
MC-Glauber model. The normalization factor of the initial entropy density
profiles are tuned to fit the multiplicity of pions in central Au+Au
collisions at $\sqrt{s_{NN}}=$ 7.7, 11.5, 19.6, 27, 39, 62.4 and 200 GeV. Since
this paper does not aim to fit the flow data at lower collision energies, we directly
set the specific shear viscosity $\eta /s=0.08$ and the specific bulk viscosity $
\zeta /s=0.0$ as once used in Ref.~\cite{Song:2010mg,Song:2012ua}. The
hydrodynamic freeze-out surface is defined by a constant
temperature, which is set to the chemical freeze-out temperature $T_{ch}$ extracted
by the statistical model at various
collision energies~\cite{Andronic:2005yp,Cleymans:2005xv}.  

This paper focuses on investigating the fluctuations of net protons.
It is thus necessary to nicely
fit the related mean values of protons and anti-protons from
the hydrodynamic calculations. Since the current
version of \texttt{VISH2+1} still inputs the equation of state s95-PCE with zero
chemical potential and does not include the transport equations for the conserved
charges, we directly add an effective chemical potential of net baryons $\mu_B$ to
the distribution function in the Cooper-Frye formula, and then fine tune $\mu_B$
to produce the yields of protons and anti-protons at each collision
energies and chosen centralities.

\begin{table}[tbp]
\caption{Parameter sets for the critical fluctuations.}
\label{parameters}\vspace{-0.15in}
\par
\begin{center}
\centering
\begin{tabular}{|p{1.7cm}|p{0.7cm}|p{0.7cm}|p{0.7cm}|p{0.7cm}|p{0.7cm}|p{0.7cm}|p{0.7cm}|}
\hline
$\sqrt{s_{NN}}[GeV]$ & ~~7.7 & ~11.5 & ~19.6 & ~~27 & ~~39 & ~62.4 & ~200 \\
\hline
\end{tabular}%
\par
\begin{tabular}{|p{1.05cm}|p{0.5cm}|p{0.7cm}|p{0.7cm}|p{0.7cm}|p{0.7cm}|p{0.7cm}|p{0.7cm}|p{0.7cm}|}
\hline
\multirow{4}*{\text{~~para-I}} & ~~$g$ & ~~3.2 & ~~2.5 & ~~2.3 & ~~2.2 & ~~~2
& ~~1.8 & ~~~1 \\ \cline{2-9}
& ~~$\tilde{\lambda}_{3}$ & ~~6 & ~~4 & ~~3 & ~~2 & ~~~0 & ~~0 &
~~~0 \\ \cline{2-9}
& ~~$\tilde{\lambda}_{4}$ & ~~14 & ~~13 & ~~12 & ~~11 & ~~10 & ~~~9 & ~~~8
\\ \cline{2-9}
& ~~$\xi $ & ~~~1 & ~~~2 & ~~~3 & ~~~3 & ~~~2 & ~~~1 & ~~0.5 \\ \hline
\end{tabular}%
\par
\begin{tabular}{|p{1.05cm}|p{0.5cm}|p{0.7cm}|p{0.7cm}|p{0.7cm}|p{0.7cm}|p{0.7cm}|p{0.7cm}|p{0.7cm}|}
\hline
\multirow{4}*{\text{~para-II}} & ~~$g$ & ~~3.2 & ~~2.5 & ~~2.3 & ~~2.2 & ~~~2
& ~~1.8 & ~~~1 \\ \cline{2-9}
& ~~$\tilde{\lambda}_{3}$ & ~~6 & ~~4 & ~~3 & ~~2 & ~~~2 & ~~1.5 &
~~~1 \\ \cline{2-9}
& ~~$\tilde{\lambda}_{4}$ & ~~14 & ~~13 & ~~12 & ~~11 & ~~10 & ~~~9 & ~~~8
\\ \cline{2-9}
& ~~$\xi $ & ~~1 & ~~2.5 & ~~4 & ~~~4 & ~~~3 & ~~~2 & ~~~1 \\ \hline
\end{tabular}%
\par
\begin{tabular}{|p{1.05cm}|p{0.5cm}|p{0.7cm}|p{0.7cm}|p{0.7cm}|p{0.7cm}|p{0.7cm}|p{0.7cm}|p{0.7cm}|}
\hline
\multirow{4}*{\text{\thinspace para-III}} & ~~$g$ & ~~2.8 & ~~1.8 & ~~1.7 &
~~1.6 & ~~~1 & ~~0.5 & ~~0.1 \\ \cline{2-9}
& ~~$\tilde{\lambda}_{3}$ & ~~6 & ~~4 & ~~3 & ~~2 & ~~~2 & ~~1.5 &
~~~1 \\ \cline{2-9}
& ~~$\tilde{\lambda}_{4}$ & ~~14 & ~~13 & ~~12 & ~~11 & ~~10 & ~~~9 & ~~~8
\\ \cline{2-9}
& ~~$\xi $ & ~~~1 & ~~~2 & ~~~3 & ~~~3 & ~~~2 & ~~~1 & ~~0.5 \\ \hline
\end{tabular}%
\end{center}
\end{table}

Besides the hydrodynamic freezeout surface, couplings $g, \ \lambda _{3},\ \lambda
_{4}$, and the correlation length $\xi$ are additional inputs in our
calculations. In order to reproduce the physical mass of protons at zero temperature,
$g\approx 10$ in the sigma model~\cite{Stephanov:2008qz}. At high
temperature, calculations from the NJL model show
that the coupling $g$ decrease as the temperature increase~\cite{Klevansky:1992qe}.
As far as we know, there is no solid calculation on how the coupling $g$
changes with temperature and chemical potential along the chemical
freeze-out line. We thus treat $g$ as a free parameter and tune it
within 0 and 10. According to the lattice simulations for the related effective
potentials around the critical point, the dimensionless parameters $\tilde{%
\lambda}_{3}\left( \lambda _{3}=\tilde{\lambda}_{3}T\left( T\xi \right)
^{-3/2}\right) $ and $\tilde{\lambda}_{4}\left( \lambda _{4}=\tilde{\lambda}
_{4}\left( T\xi \right) ^{-1}\right) $ separately ranges between $(0,8)$ and
$(4,20)$ from crossover to the first order phase transition~\cite{Tsypin:1994be, Stephanov:2008qz}.
Considering the critical slowing down near the critical point, the maximum value
of the correlation length $\xi$ is set to O(3fm)~\cite{Berdnikov:1999ph,Nonaka:2004pg}.
Away from the critical point, the correlation length $\xi$ gradually decreases
to its natural value around 0.5-1 fm~\cite{Stephanov:2008qz,Stephanov:2011pb}.

Table I lists three parameter sets used in our calculations. Here, $g, \ \tilde{\lambda} _{3},\ \tilde{\lambda}
_{4}$ and $\xi$ are tuned within the above constrains to roughly fit $C_4$
and to describe the decreasing trend of $C_2$ and $C_3$ with Poisson and Binomial
baselines. Note that these parameter
sets are not unique since the measured cumulants $C_2$ and $C_3$ are always
below the Poisson/Binomial baselines, which can not strictly
constrain the related parameters of critical fluctuations (please refer to Sec.~IV for
details).

In order to compare with the experimental data, one also needs to consider
the contributions from trivial statistical fluctuations (noncritical fluctuations).
Here, we assume that the critical and noncritical
fluctuations are independent, which can be treated separately. Following
Ref.~\cite{Luo:2014tga}, we take either Poisson or Binomial
distributions as the trivial statistical fluctuations. If protons and
anti-protons are independently produced and satisfy the Poisson distributions,
the net protons satisfy the Skellam distributions. The correspondent
cumulants and cumulants ratios are expressed as: $%
C_{1}=C_{3}=M_{p}-M_{\bar{p}}$, $C_{2}=C_{4}=M_{p}+M_{\bar{p}}$, $S\sigma =%
\frac{C_{3}}{C_{2}}=\frac{M_{p}-M_{\bar{p}}}{M_{p}+M_{\bar{p}}}$, $\kappa
\sigma ^{2}=\frac{C_{4}}{C_{2}}=1$, where $M_{p}$ and $M_{\bar{p}}$ are the
mean values of protons and antiprotons. If protons and anti-protons satisfy
independent Binomial distributions, the cumulants of protons and
anti-protons are written as: $C_{1}^{x}=M_{x}$, $C_{2}^{x}=\varepsilon _{x}M_{x}$,
$C_{3}^{x}=\varepsilon _{x}M_{x}\left( 2\varepsilon _{x}-1\right) $,
$C_{4}^{x}=\varepsilon _{x}M_{x}\left( 6\varepsilon _{x}^{2}-6\varepsilon
_{x}-1\right) $, where $M_{x}$ is the mean value of protons/anti-protons and $\varepsilon _{x}$
is an additional parameter, which is determined by $\varepsilon _{x}=C_{2}^{x}/M_{x}$. Then, the
cumulants of the net-protons are expressed as: $%
C_{n}^{net-p}=C_{n}^{p}+\left( -1\right) ^{n}C_{n}^{\bar{p}}$~\footnote{As pointed in
Ref.~\cite{Kapusta:2011gt,Young:2014pka}, a hydrodynamic system has
imprinted thermal fluctuations according to the fluctuation-dissipation theorem.
Considering the numerical complexities, we do not further
explore such thermal fluctuations, but only take the Poisson or Binomial
distributions as the trivial statistical fluctuations. }.

\section{Numerical results}.

\begin{figure*}[tbp]
\centering
{\normalsize \textbf{Au+Au 0-5\%, \ noncritical + critical fluctuations (Poisson
baselines)} }
\includegraphics[width=3.5 in,height=3.3 in]{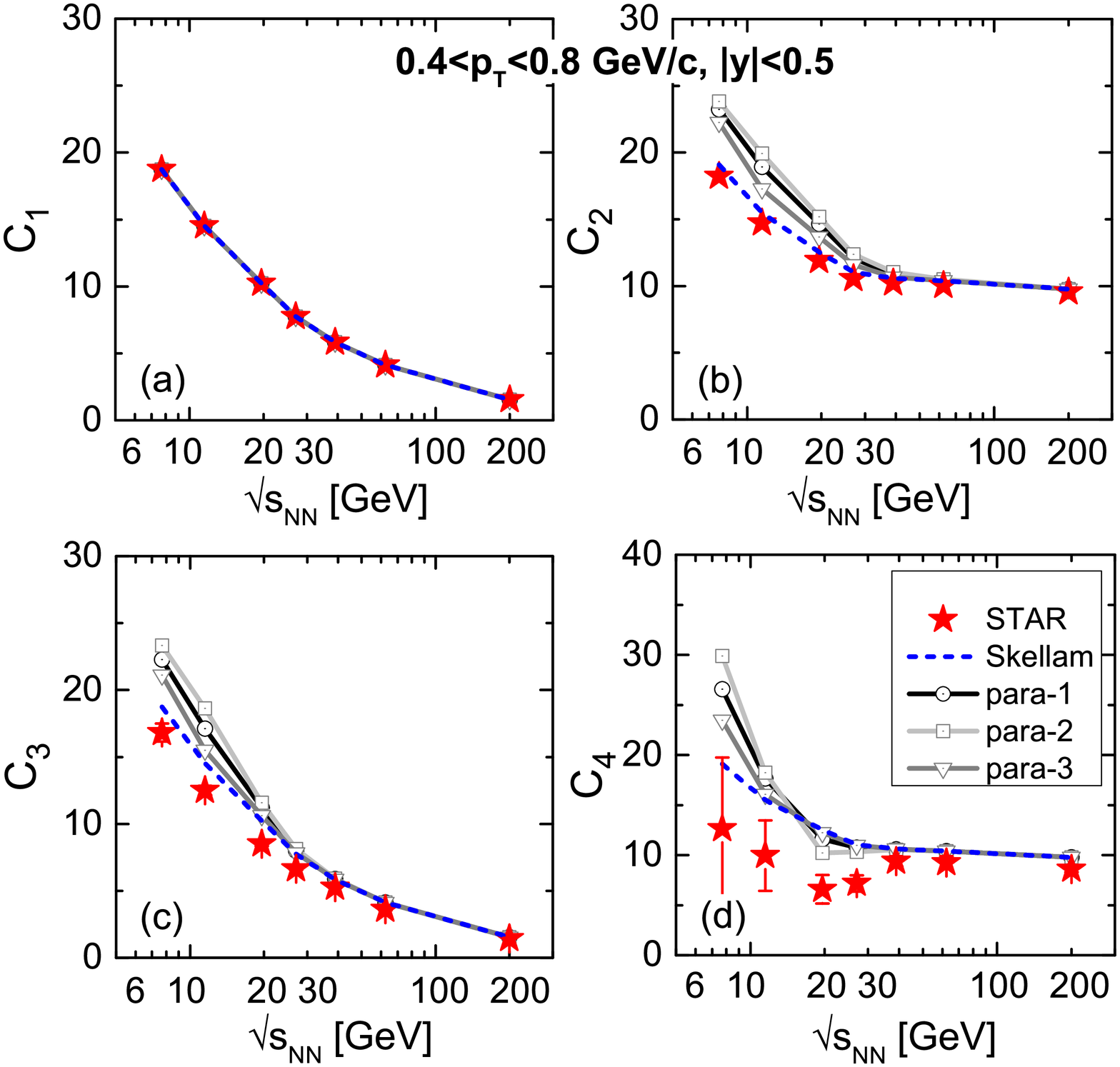} %
\includegraphics[width=3.5 in,height=3.3 in]{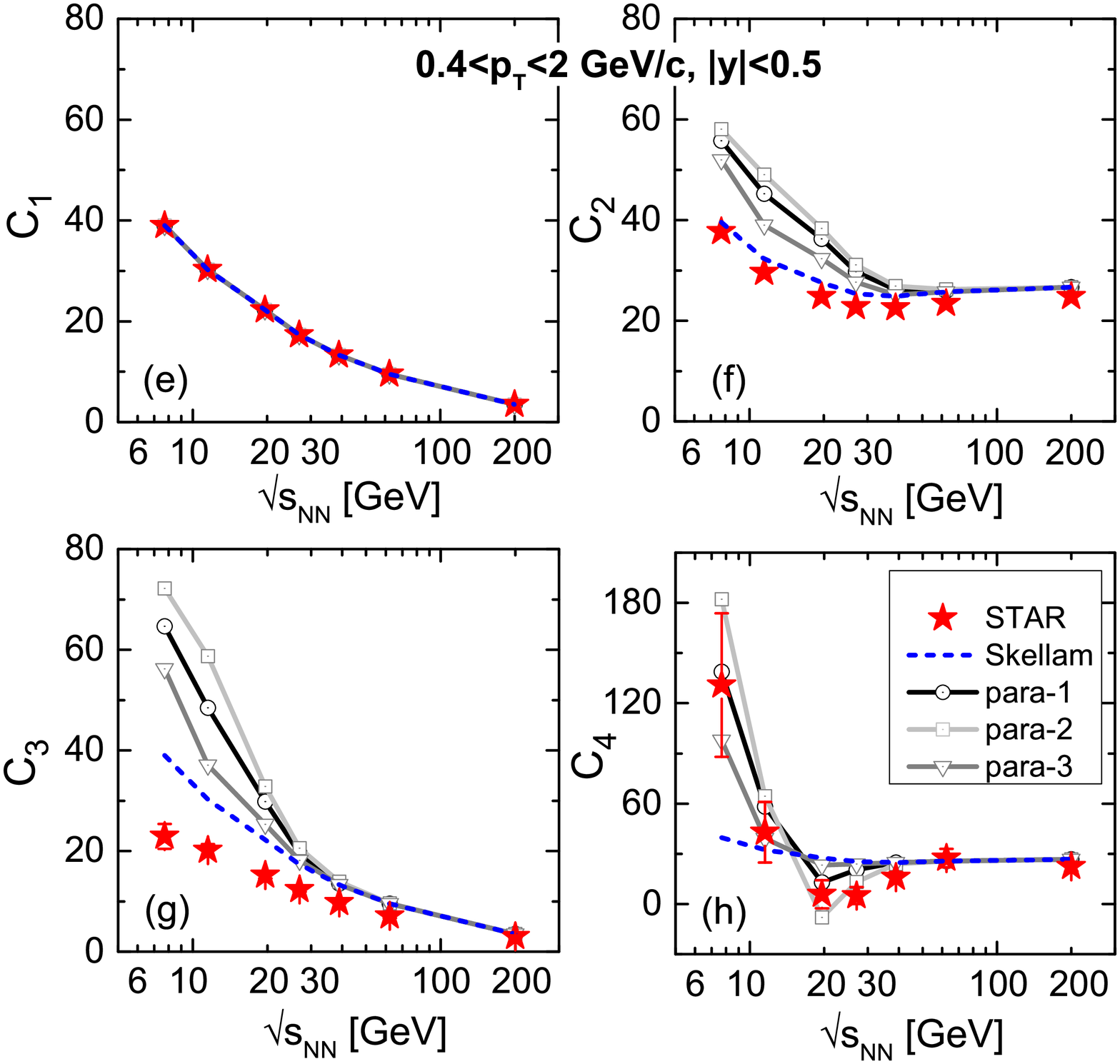} 
\caption{(Color online) Energy dependence of cumulants $C_1\sim C_4$  for net protons in 0-5\%  Au+Au collisions within $0.4 <p_T< 0.8\ \mathrm{GeV}$ (left panels) and within $0.4 <p_T< 2\ \mathrm{GeV}$ (right panels).  The red stars are the STAR preliminary data~\cite{Adamczyk:2013dal,Luo:2015ewa}, dashed blue lines are the Poisson expectations, and black and grey curves with symbols are the results from our model calculations with the Poisson baselines.}
\label{cumulant4-ske-0005}
\end{figure*}
\begin{figure*}[tp]
\centering
{\normalsize \textbf{Au+Au 0-5\%, \ noncritical + critical fluctuations (Binomial
baselines)} }
\includegraphics[width=3.5 in,height=3.3 in]{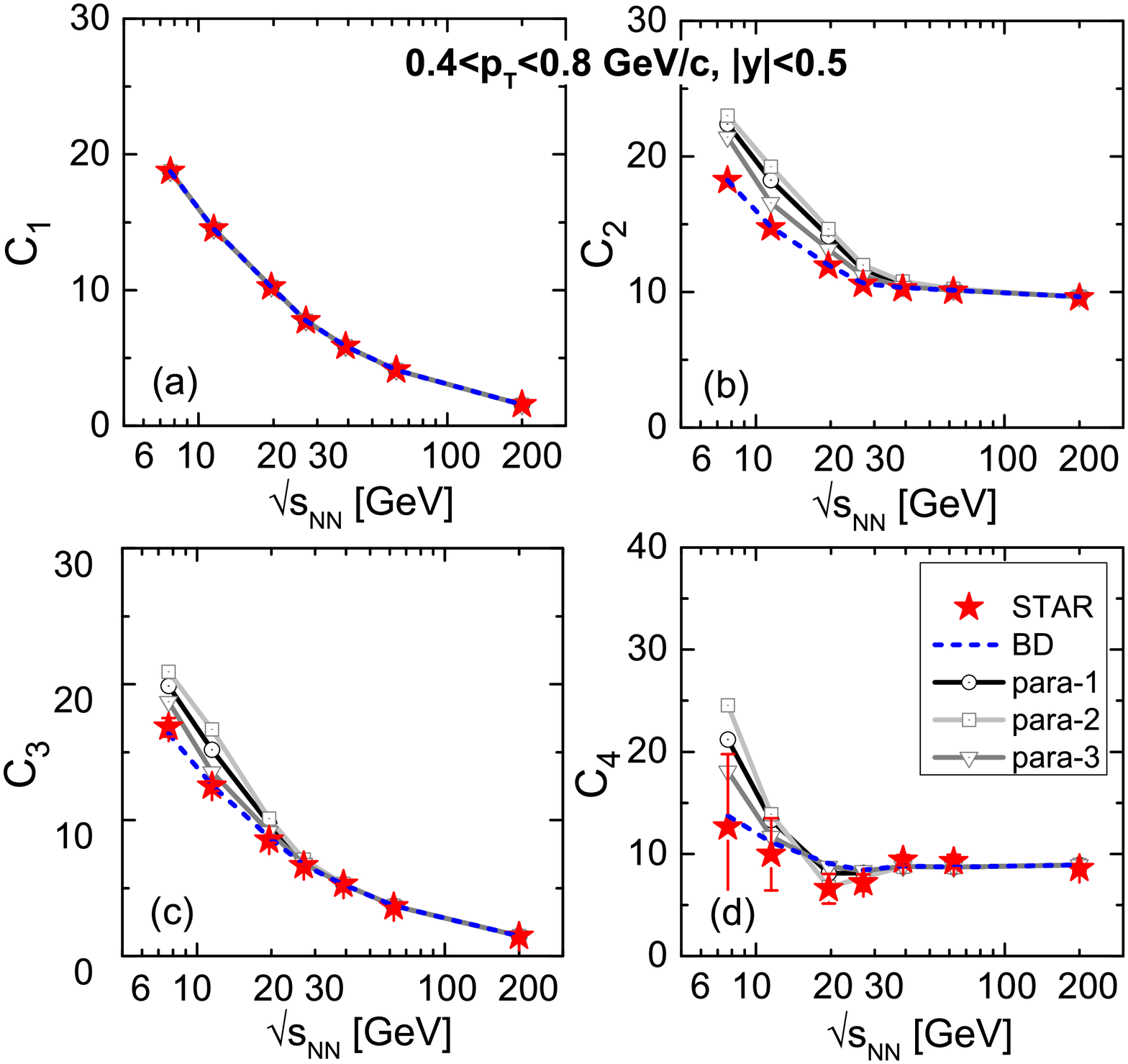} %
\includegraphics[width=3.5 in,height=3.3 in]{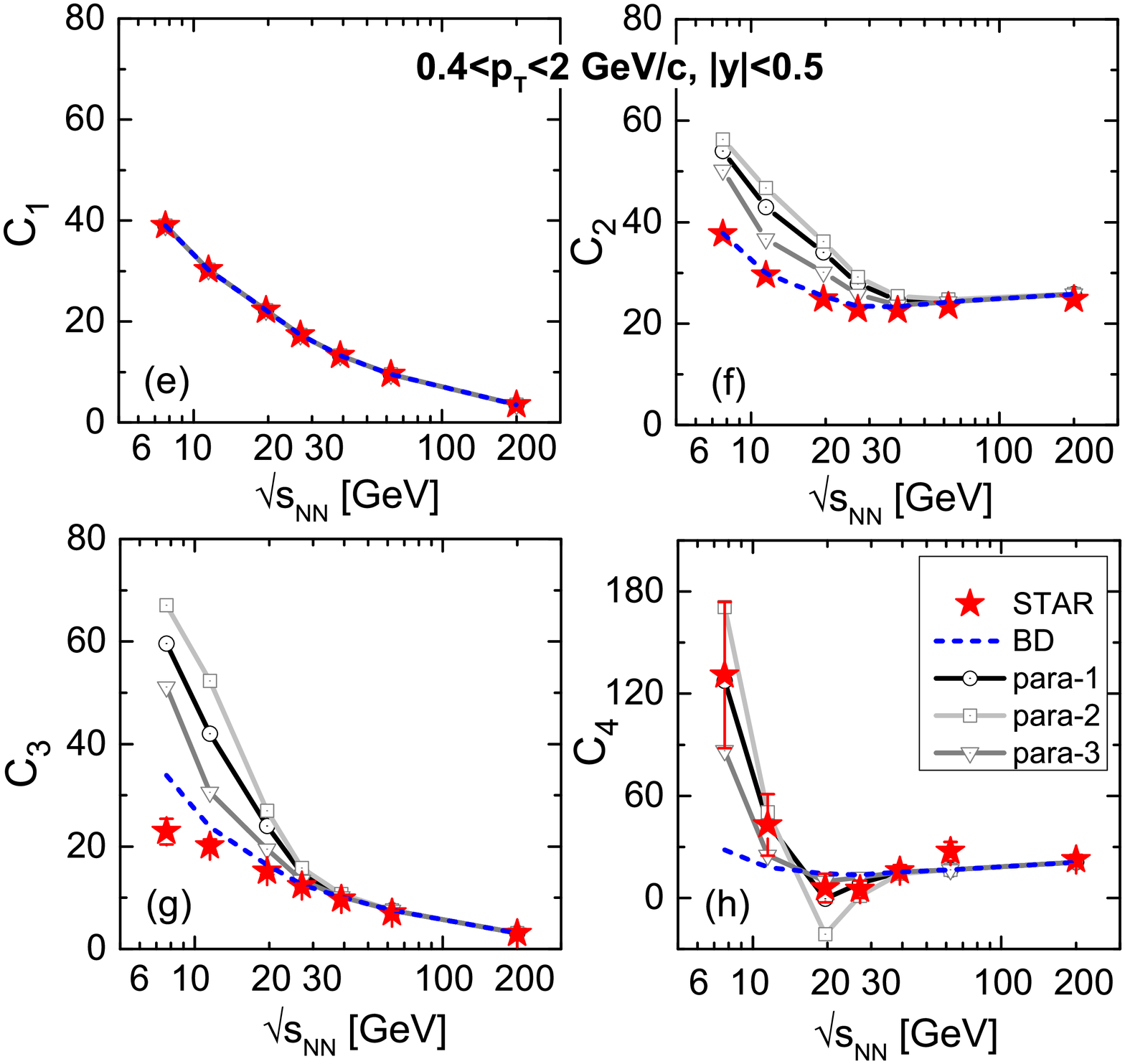} 
\caption{(Color online) Similar to Fig.1, but with Binomial baselines.}
\label{cumulant4-BD-0005}
\end{figure*}

\begin{figure*}[tbp]
\centering
{\normalsize \textbf{Au+Au 30-40\%, \ noncritical + critical fluctuations (Poisson baselines)} }
\includegraphics[width=3.5 in,height=3.3 in]{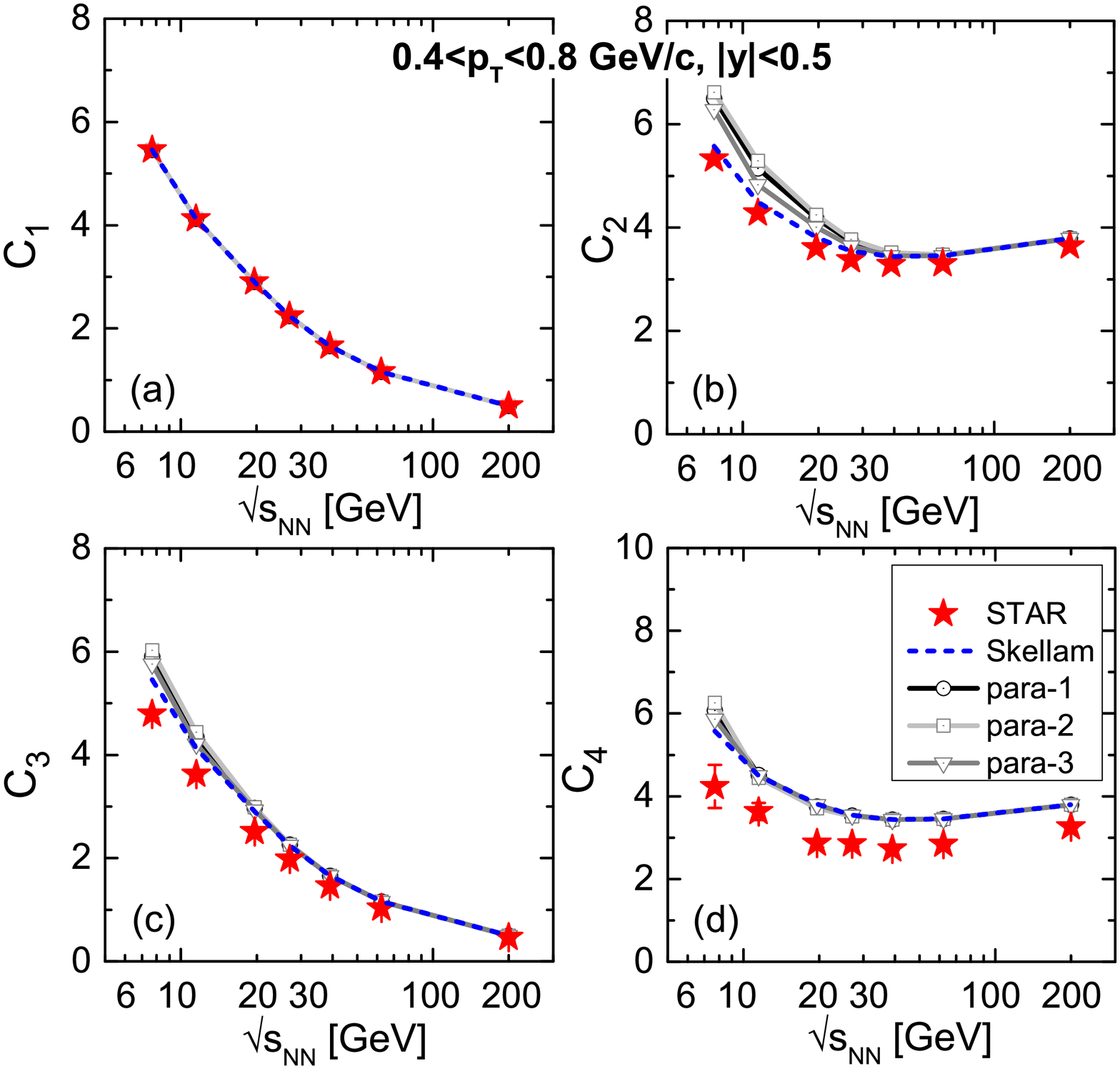} %
\includegraphics[width=3.5 in,height=3.3 in]{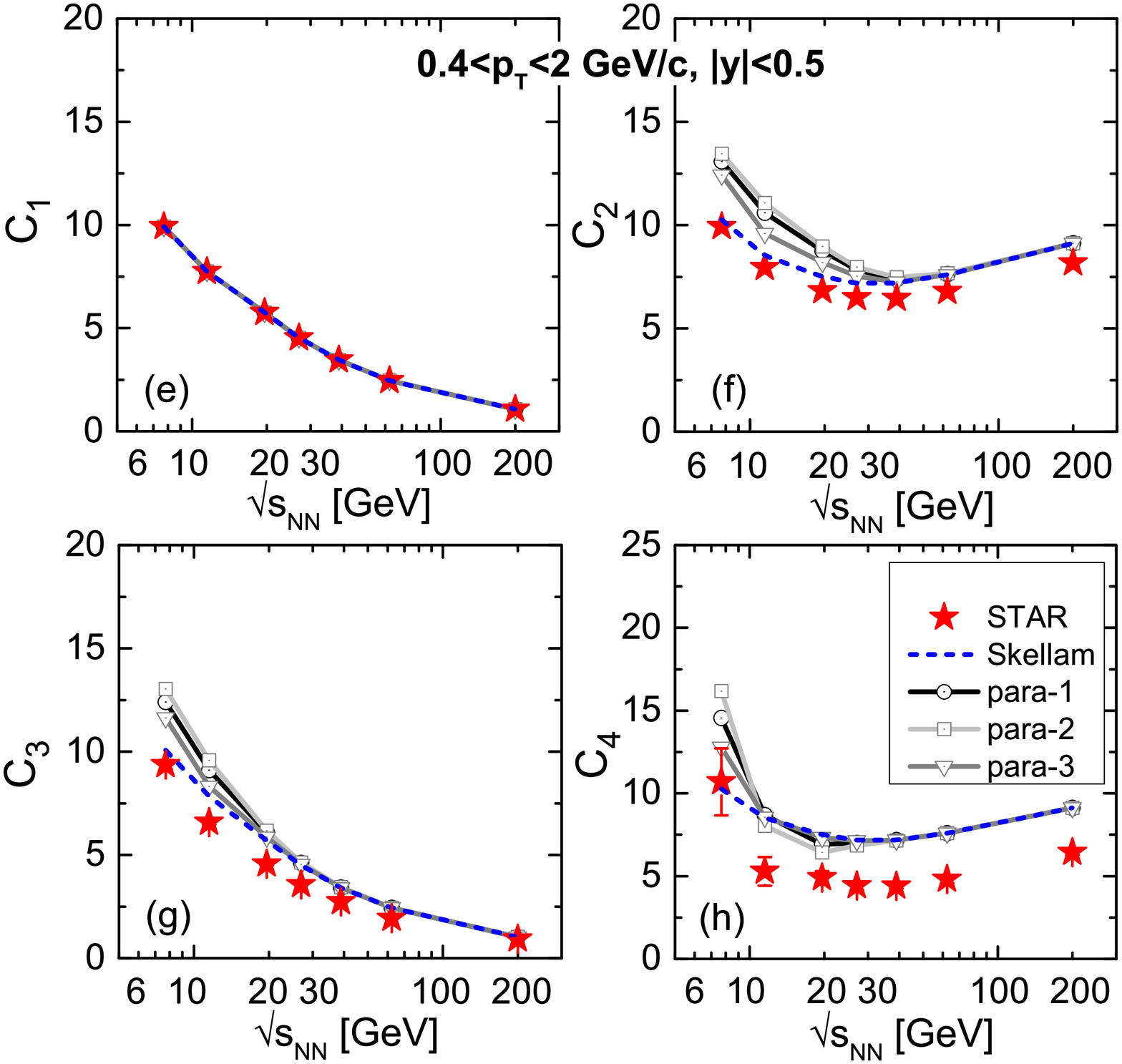}
\caption{(Color online) Energy dependence of cumulants $C_1\sim C_4$ for net protons in 30-40\% Au+Au collisions, with Poisson baselines.}
\label{cumulant4-ske-3040}
\end{figure*}

\begin{figure*}[tp]
\centering
{\normalsize \textbf{Au+Au 30-40\%, \ noncritical + critical fluctuations
(Binomial baselines)} }
\includegraphics[width=3.5 in,height=3.3 in]{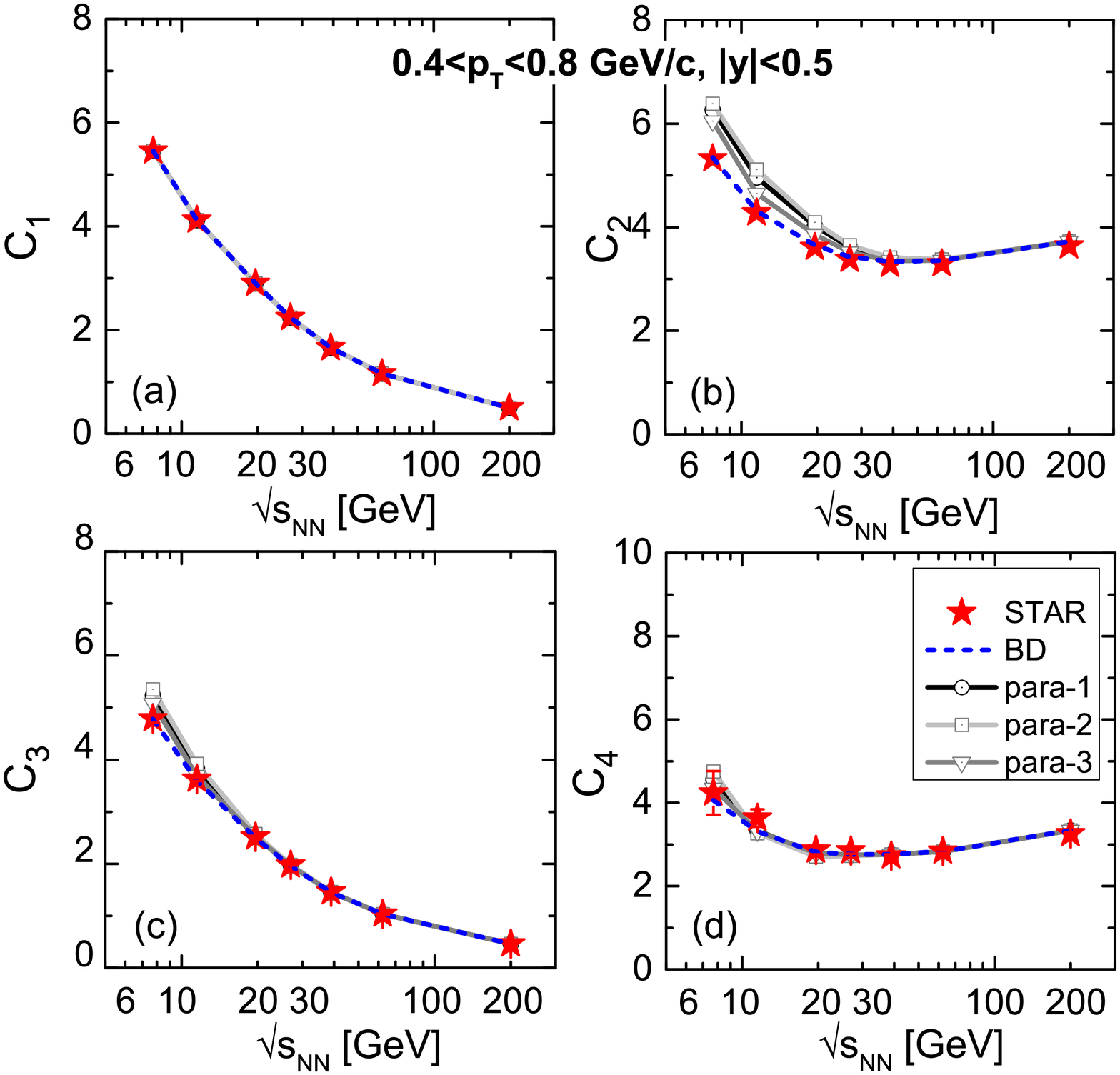} %
\includegraphics[width=3.5 in,height=3.3 in]{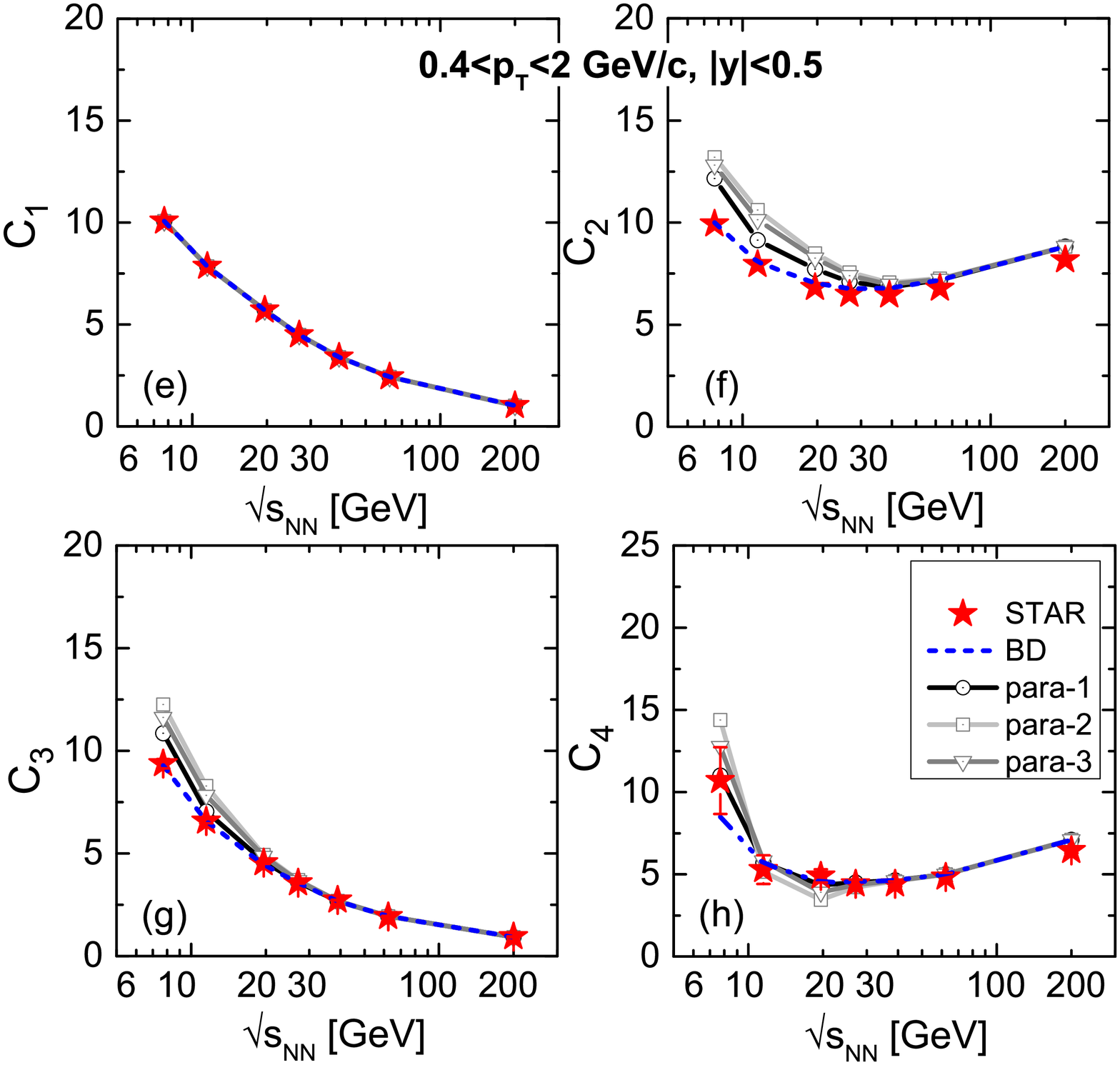} 
\caption{(Color online) Similar to Fig. 3, but with Binomial baselines.}
\label{cumulant4-BD-3040}
\end{figure*}

\begin{figure*}[t]
\centering
\includegraphics[width=3.3 in]{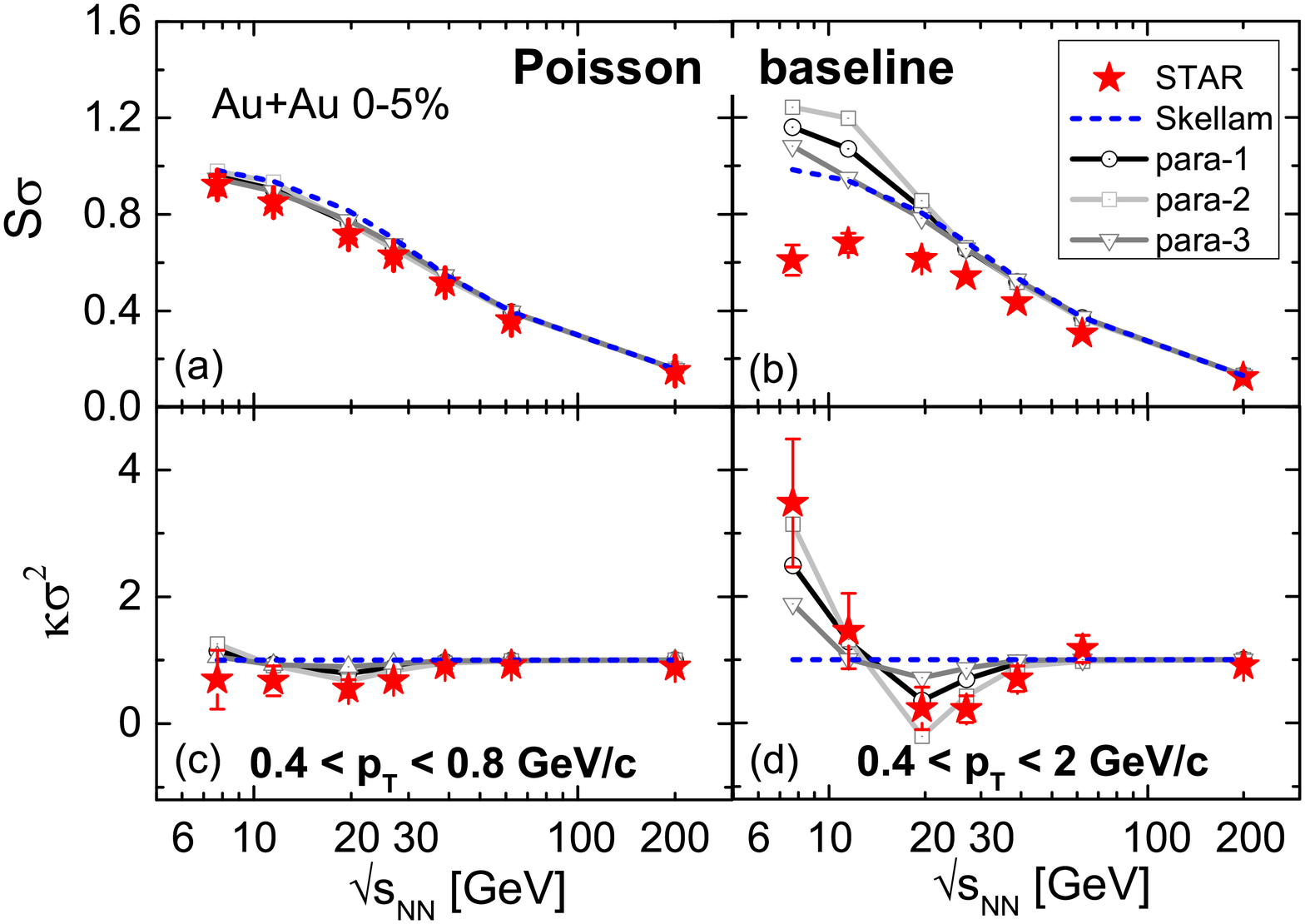} %
\includegraphics[width=3.3 in]{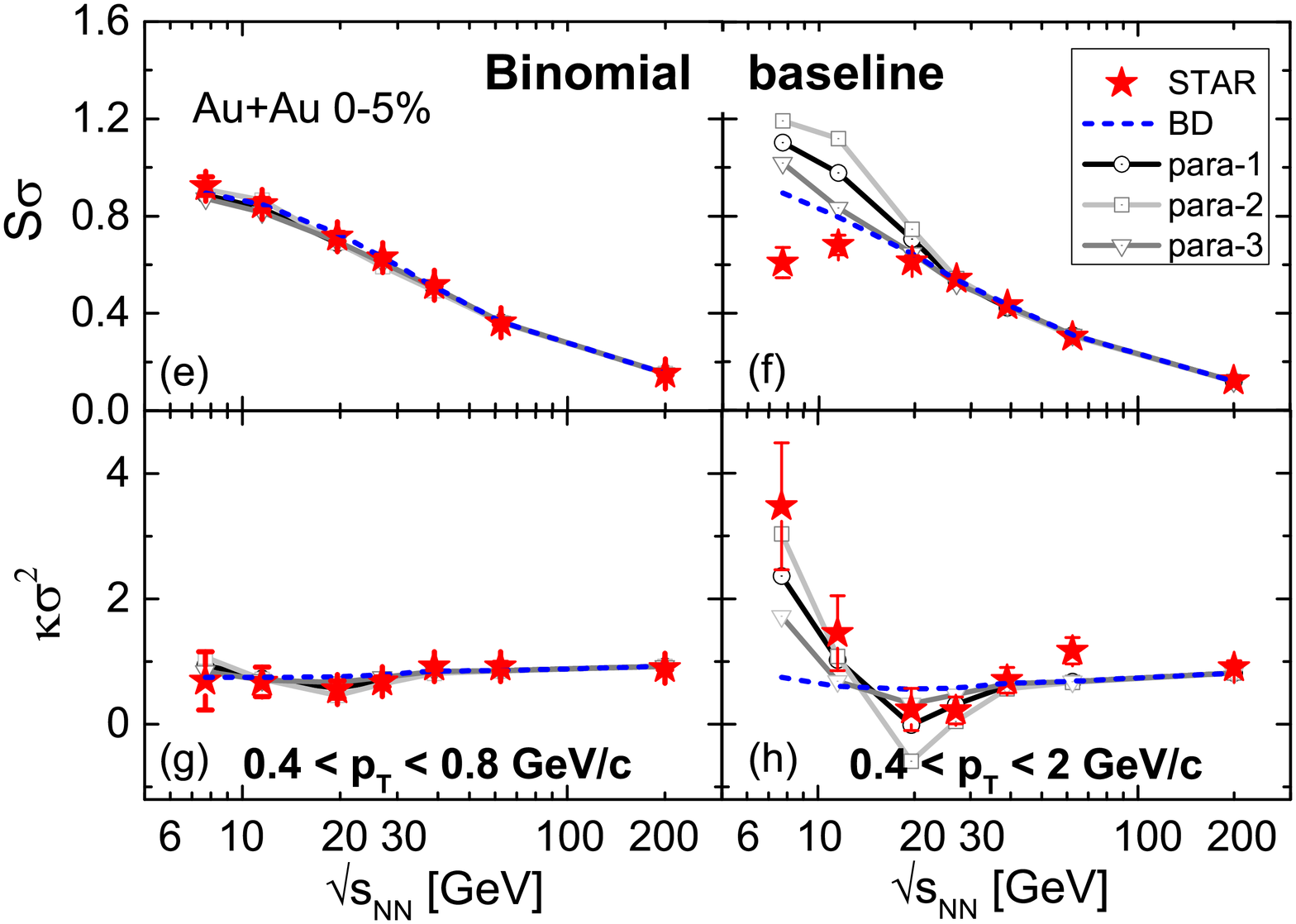}
\caption{(Color online) Energy dependence of cumulant ratios, $S \sigma$ and $\kappa \sigma^2$, for net protons in  0-5\%
Au+Au collisions, with different $p_{T}$ ranges and different non-critical fluctuation baselines.}
\label{c-ratio-1}
\end{figure*}
\begin{figure*}[tbp]
\centering
\includegraphics[width=3.3 in]{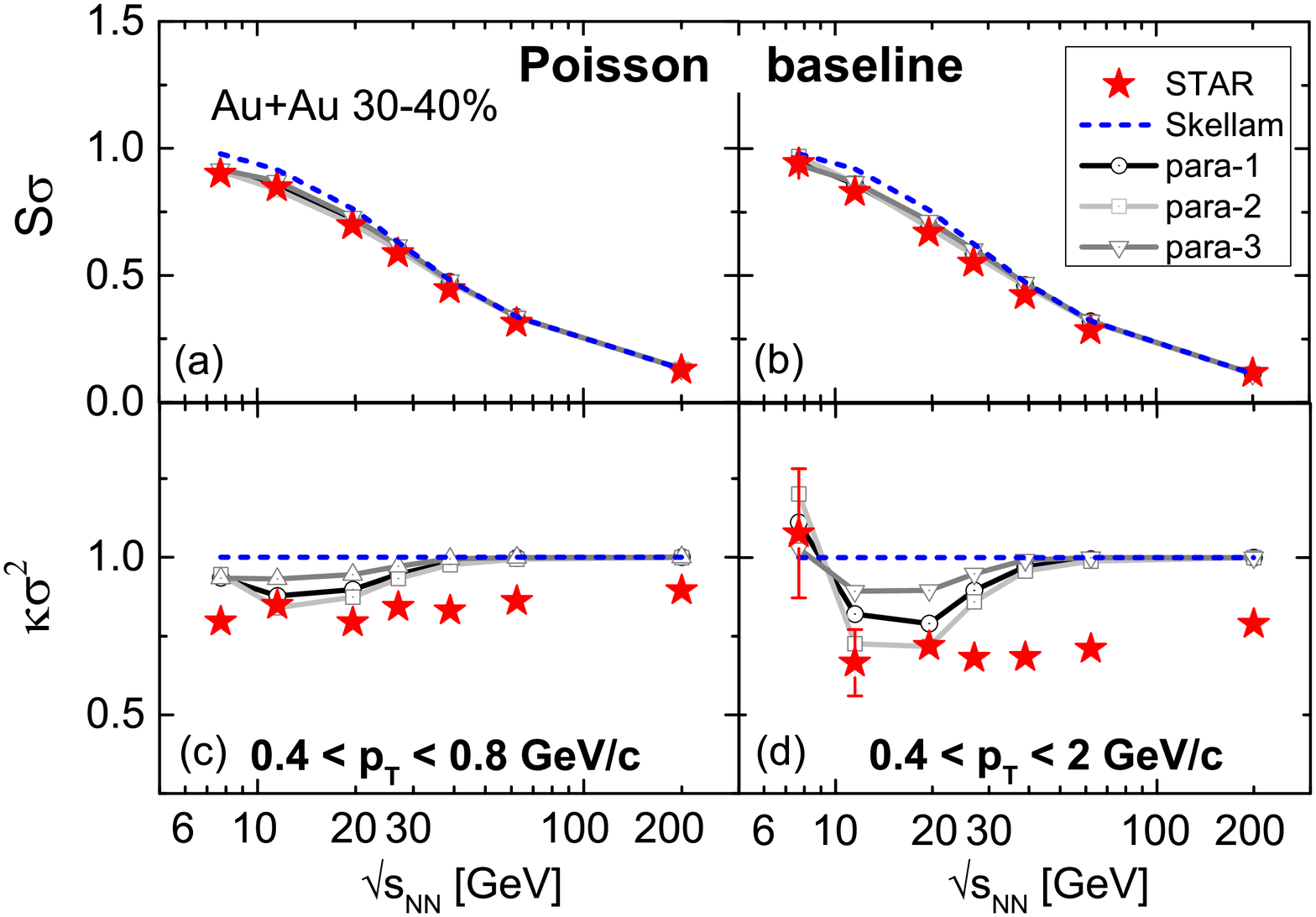} %
\includegraphics[width=3.3 in]{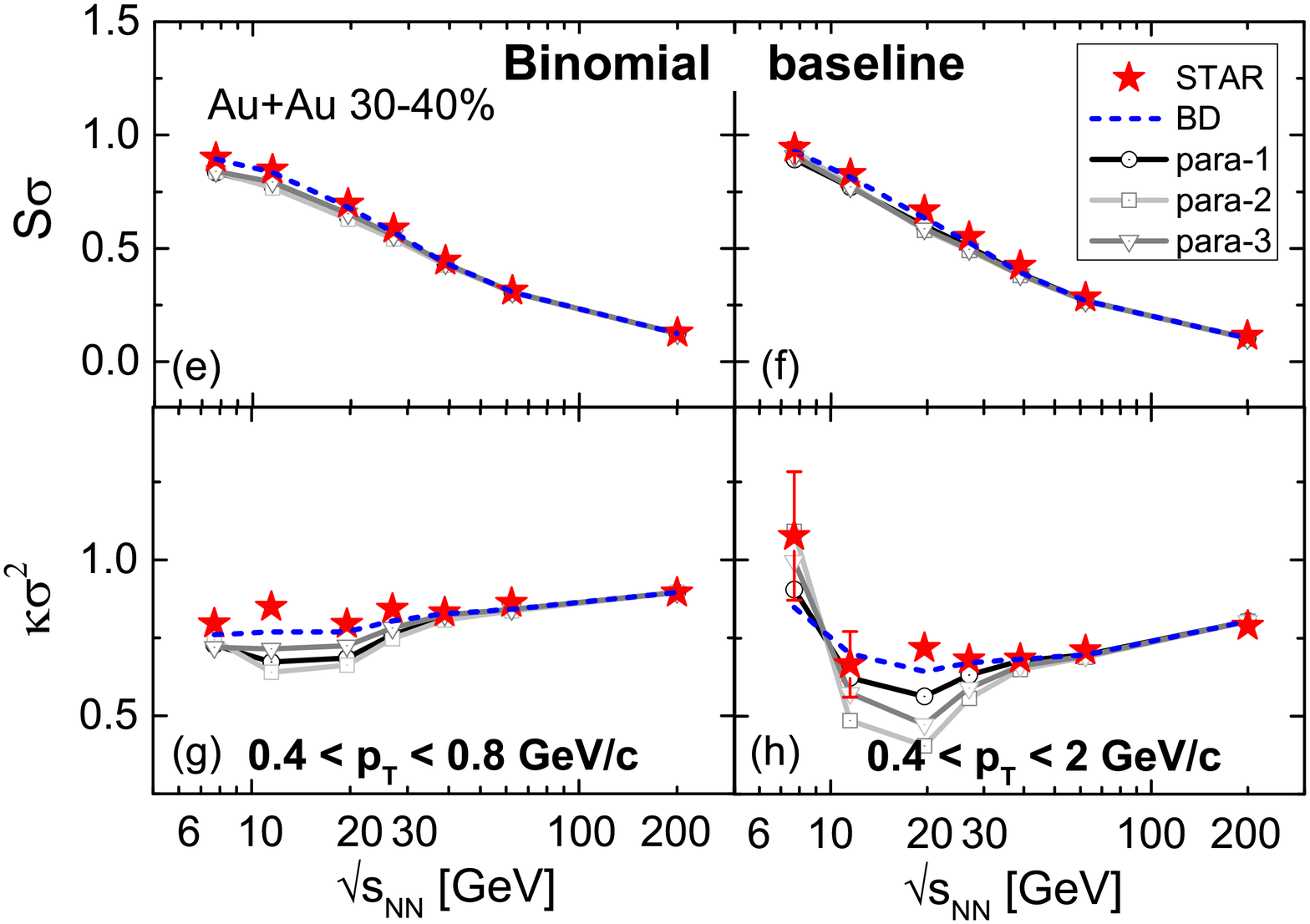} 
\caption{(Color online) Similar to Fig.~5, but for 30-40\% centrality bins.}
\label{c-ratio-2}
\end{figure*}

With the formulism and set-ups presented in Sec.II and Sec.III,
we calculate the correlated fluctuations of net protons emitted
from the hydrodynamic freeze-out surface with the presence of an
order parameter field. Fig.~1 and Fig.~2 show the energy dependent
cumulants $C_{1}\sim C_{4}$ of net protons in central Au+Au collisions
with either Poisson or Binomial distributions served as the trivial statistical
fluctuation baselines, where the left and right panels
present the results within two different
transverse momentum ranges, $0.4<p_{T}<0.8 \ \mathrm{GeV} $ and $0.4<p_{T}<2 \ \mathrm{GeV}$, respectively.
The red stars are the STAR preliminary data~\cite{Adamczyk:2013dal,Luo:2015ewa},
the dashed blue lines are the expectation
values of Poisson/Binomial distributions, and the black and grey curves with
different symbols are the theoretical results from our model calculations.
Here, we implement three parameters sets, as listed in table I, to calculate the
the critical fluctuations~\footnote{The critical fluctuations are calculated
through integrating the static correlators of $\delta f$ over the whole freeze-out surface.
If constraining the correlations within certain time slice, e.g. $\Delta t\leq 1 \mathrm{fm}$,
$C_2^{critical}$, $C_3^{critical}$ and $C_4^{critical}$  at lower collision energies  would respectively reduce by O(10\%), O(30\%) and O(50\%)
with the same parameter sets as inputs. After re-tuning the related parameters, similar results
as shown in Fig.~1 and Fig.~2 could be obtained even with such constraint.
}.
With the assumption that non-critical and critical
fluctuations are independent, the total cumulants $C_2\sim C_4$ (black and grey curves)
are obtained by summing up the cumulants from the critical fluctuations and
from the Poisson/Binomial fluctuations.

In our model calculations, $C_{1}$, the mean value of net protons, is used
to set the chemical potential of net baryons in the Cooper-fryer formula, which
leads to a nice fit of the $C_1$ data within different $p_T$ ranges
at various collision energies. $C_{2}\sim C_{4}$
are the cumulants to evaluate fluctuations. Generally, the deviations between
the data and the Poisson/Binomial baselines are considered
as the contributions from the critical fluctuations. After tuning $g$, $\xi$, $\tilde{\lambda}_{3}$
and $\tilde{\lambda}_{4}$ within the allowed parameter ranges, we could roughly
describe the decrease trend of $C_2$ and $C_3$ and the non-monotonic
behavior of $C_4$ with the increase of collision energy.
However, $C_{2}$ and $C_{3}$ from our model calculations are always above the
Poission/Binomial baselines due to the positive contributions from the static
critical fluctuations, which shows even larger deviations from the experimental data.
Similarly, the standard
critical fluctuations from Stephanov in a static and infinite medium
also give positive contributions to $C_2$ and $C_3$. Such sign problem of $C_2^{critical}$
and $C_3^{critical}$
might be solved if further considering the time evolution for the correlated
fluctuating order parameter field, which should be studied in the future.

For the energy dependent cumulant $C_{4}$, our model calculations can roughly describe the experimental data
within both $0.4<p_{T}<0.8 \ \mathrm{GeV}$ and $0.4<p_{T}<2 \ \mathrm{GeV}$ with the Binomial
baselines. However, if using the Poisson baselines,
it's difficult to simultaneously describe $C_{4}$
within these two different $p_{T}$ ranges for Au+Au collisions at lower collision energies.
At small collision energies below 11.5 GeV, the measured $C_{4}$ are higher than the poisson expectation
values for $0.4<p_{T}<2 \ \mathrm{GeV}$, but lower than the poisson
expectation values for $0.4<p_{T}<0.8 \ \mathrm{GeV}$. According to Eqs.(13-15), the change of the $p_{T}$
ranges only affects the magnitude of $C_{n}^{critial}$ of the critical fluctuations, rather than
their signs. As a result, our calculations with the Poisson baselines can not simultaneously fit the $C_{4}$
data within these two $p_T$ ranges.

Fig.~1 and Fig.~2 also show that, with the $p_{T}$ range changed from $%
0.4<p_{T}<0.8 \ \mathrm{GeV}$ to $0.4<p_{T}<2 \ \mathrm{GeV}$, the measured
$C_{4}$ dramatically increases, showing much larger deviations from
the Poison/Binomial baselines. It is thus generally believed that large
acceptance is a crucial factor to probe the critical fluctuations in experiment.
In our model calculations, the $n_{th}$ cumulants of
the critical fluctuations is closely related to the $n_{th}$ power of the
total net-proton number within the defined $p_T$ and rapidity range. With
the maximum $p_T$ increased from $0.8 \ \mathrm{GeV}$ to $2 \ \mathrm{GeV}$,
the averaged number (mean value) of net protons almost increases by a
factor of two, leading to a dramatic increase of $C_{3}^{critical}$ and $%
C_{4}^{critical}$ in our calculations.  One may also notice that, although
the correlation length $\xi$ decreased to 1 fm in Au+Au collisions at
$\sqrt{s_{NN}}=$7.7 GeV, the critical fluctuations of $C_2$, $C_3$ and
$C_4$ are still large there, especially for the cases with
$0.4<p_{T}<2 \ \mathrm{GeV}$. This is mainly because the net proton numbers within specific
acceptance range dramatically increase at lower collision energies. Compared with the case at
$\sqrt{s_{NN}}=$19.6 GeV, the mean values of net protons $C_1$ within both $p_T$ ranges
almost increase by a factor of 2 for $\sqrt{s_{NN}}=$7.7 GeV, leading to larger $C_{2}^{critical}$, $C_{3}^{critical}$ and $%
C_{4}^{critical}$ even with $\xi$ = 1 fm.

Similarly, the higher cumulants of critical fluctuations are dramatically suppressed from
central to semi-cental collisions due to the decreased mean
values of net protons. To further explore such effects, we extend our
calculations to 30-40\% centrality bin with the same inputs and
parameter sets. Fig.~3 and Fig.~4 show the energy dependent cumulants
$C_{1}\sim C_{4}$ in 30-40\% Au+Au collisions
with Poisson and Binomial baselines, respectively.
Due to largely reduced critical
fluctuations, these black and grey curves from our model calculations are
close to the Binomial and Poisson baselines.  We have also noticed that the Poisson
expectation values are well above the $C_{4}$ data for different $p_T$
ranges, which make it impossible to fit the $C_{4}$ if using the
Poisson baselines in our model calculations.
In contrast, the Binomial baselines are very close to the measured
$C_{1}\sim C_{4}$ in 30-40\% centrality, with which we can roughly fit the $C_{4}$
data within error bars. However, $C_{2}$ and $C_{3}$ are still
slightly over-predicted there due to the positive contributions from
the critical fluctuations.

Fig.~5 and Fig.~6 show the energy dependent cumulant ratios $S\sigma
=C_{3}/C_{2}$ and $\kappa \sigma ^{2}=C_{4}/C_{2}$ in 0-5\% and 30-40\%
Au+Au collisions, where the left and right panels present the
results with Poisson and Binomial baselines, respectively. Although $C_{2}$ and $C_{3}$
are over-predicted in our model calculations, the cumulant ratios $S\sigma $ and $%
\kappa \sigma ^{2}$ show better agreement with the experimental data
in central Au+Au collisions, except for $S\sigma $ within
$0.4<p_{T}<2 \ \mathrm{GeV}$ (pannel (b) and (f) in Fig.~5). For
non-central Au+Au collisions at $30-40\%$ centrality, the Binomial baselines are very close to the experimental data. Together with
the dramatically reduced critical fluctuations, this
leads to a nice fit of the experimental data in our model calculations.
In contrast, the Poisson baselines are well above the measured $\kappa \sigma ^{2}$
for almost all collision energies at the 30-40\% centrality bin.  Due to  small contributions
from the critical fluctuations, it is impossible to fit $\kappa \sigma ^{2}
$ at higher collision energies above 39 GeV in our model calculations with the Poisson baselines.

\section{Summary and discussions}.

In this paper, we introduced a freeze-out scheme for the dynamical models near the QCD critical
point through coupling the decoupled classical particles with the correlated fluctuating sigma field. With a modified distribution
function that satisfies specific static fluctuations, we deduced the formulism to calculate
the 2nd, 3rd and 4th cumulants for the particles emitted from the hydrodynamic
freeze-out surface with the presence of an order parameter field.  We proved that, for a static
and infinite medium, this formulism can reproduce the early results of static fluctuations deduced by Stephanov in 2009.

With the parameters tuned within the allowed ranges, we calculated the correlated
fluctuations of net protons on the hydrodynamic freeze-out surface in
central and non-central Au+Au collisions at various collision energies.
With Poisson or Binomial baselines,
our model calculations could roughly describe the decrease trend of the cumulants $C_2$ and $C_3$
and the non-monotonic behavior of the cumulant $C_4$ as observed in experiment, but always over-predict
the values of $C_{2}$ and $C_{3}$ due to the positive contributions from
the static critical fluctuations.

After fine tuning the coupling $g$, $\xi$, $\lambda_3$ and $\lambda_4$, our model calculations,
with the Binomial baselines, can roughly describe the energy dependence of  $C_4$ and
$\kappa \sigma ^{2}$ within different $p_T$ ranges in both central and non
central collisions. However, if using the Poisson baselines, it is difficult to simultaneously describe
$C_4$ within different $p_T$ ranges in central Au+Au collisions at
lower collision energies. Meanwhile, it is impossible to fit $C_4$ and
$\kappa \sigma ^{2}$ with the much reduced critical fluctuations for Au+Au collisions
at 30-40\% centrality bin since the Poisson baselines largely deviate from the experimental data there.
For our model calculations, the Binomial distributions with two parameters
seem to give better non-critical fluctuations baselines when compared with the one-parameter Poisson distributions.
However, how the trivial statistical fluctuations, like Poisson distributions,
are destroyed through the effects like hadronic scattering and decays,
the cut of centrality bins, etc. has not been fully investigated,
which should be studied in the near future.

At last, we emphasize that our calculations presented in this paper belong to the
category of static critical fluctuations due to the static correlators of the
sigma field used in our formulism. In order to qualitatively
and quantitatively describe the experimental data, especially to solve the sign problem
of $C_2$ and $C_3$ for the critical fluctuations,
the dynamical evolution of the order parameter field and other related effects should be further investigated.
Along this direction, Mukherjee and his collaborators have done some pioneering
work on solving the evolution equations of various cumulants for the sigma field
and found that Skewness and Kurtosis could change their sign after
the dynamical evolution~\cite{Mukherjee:2015swa}. However, their approach can not be directly
implemented in our formulism on the freeze-out surface since only zero mode of sigma field are
considered in their calculations, which already erase the needed spacial information.
Dynamical models near the critical point, eg. chiral hydrodynamics, involve a full evolution
of the sigma field in position space, together with the coupled evolution of the bulk matter~\cite{Paech:2003fe,Nahrgang:2011vn,Nahrgang:2011mg,Herold:2013bi}.
Combined with the freeze-out scheme introduced in this paper,
one could, in principle, calculate the correlated fluctuations of the classical
particles decoupled from the bulk matter with the presence of an evolving order parameter field.
However, a quantitative study of the higher cumulants of produced hadrons also requires
sophisticated simulations for the evolution equation of the sigma field with the
properly chosen noise and dissipation terms and initial conditions with correlated fluctuations,
etc.. Part of the related work have been done
by different groups in the past few years~\cite{Paech:2003fe,Nahrgang:2011vn,
Nahrgang:2011mg,Herold:2013bi}, more progresses are expected in the near future.

\section*{Appendix}.

\begin{figure*}[tbp]
\centering
\includegraphics[width=3.3 in,height=3.2 in]{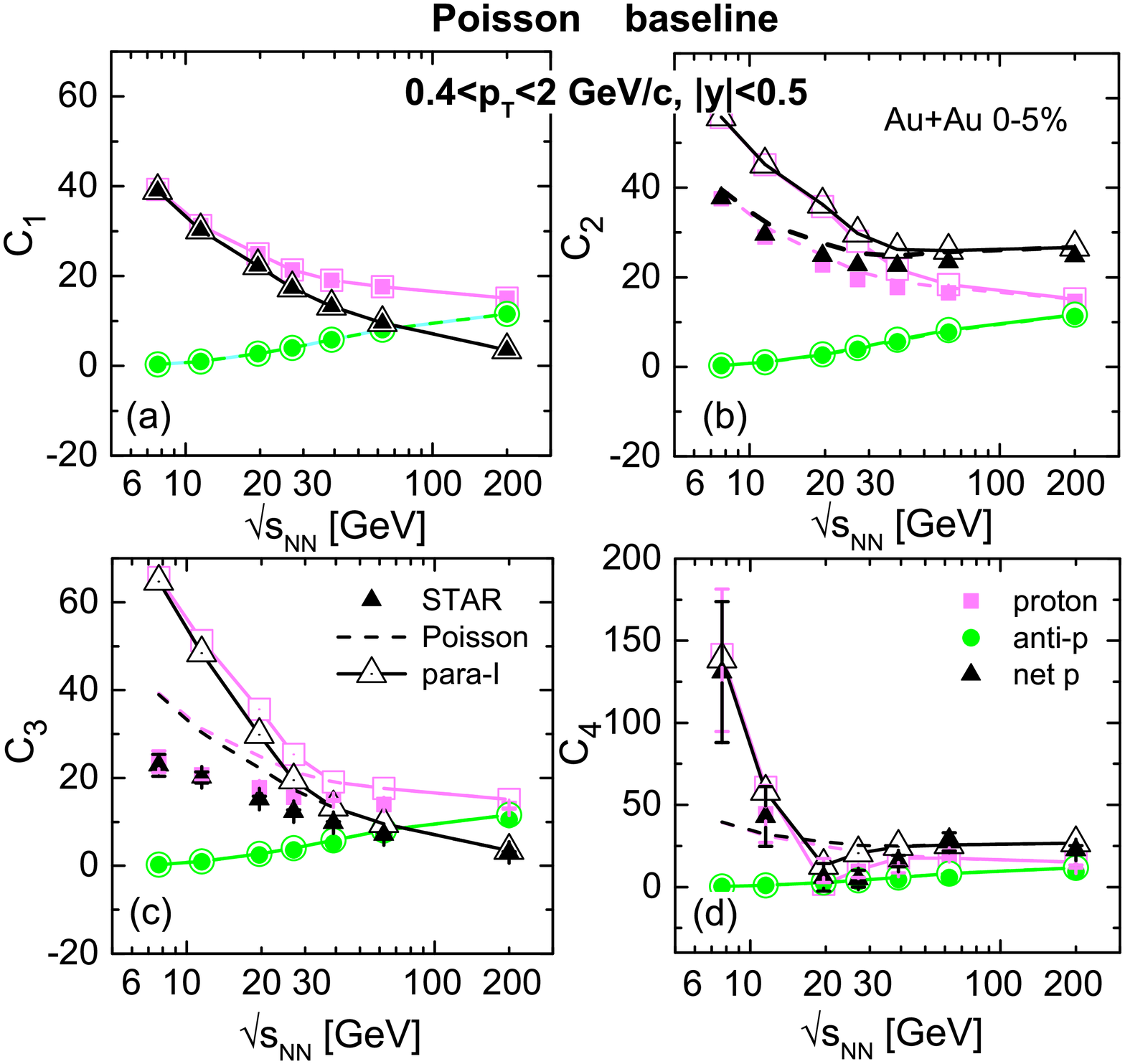} %
\includegraphics[width=3.3 in,height=3.2 in]{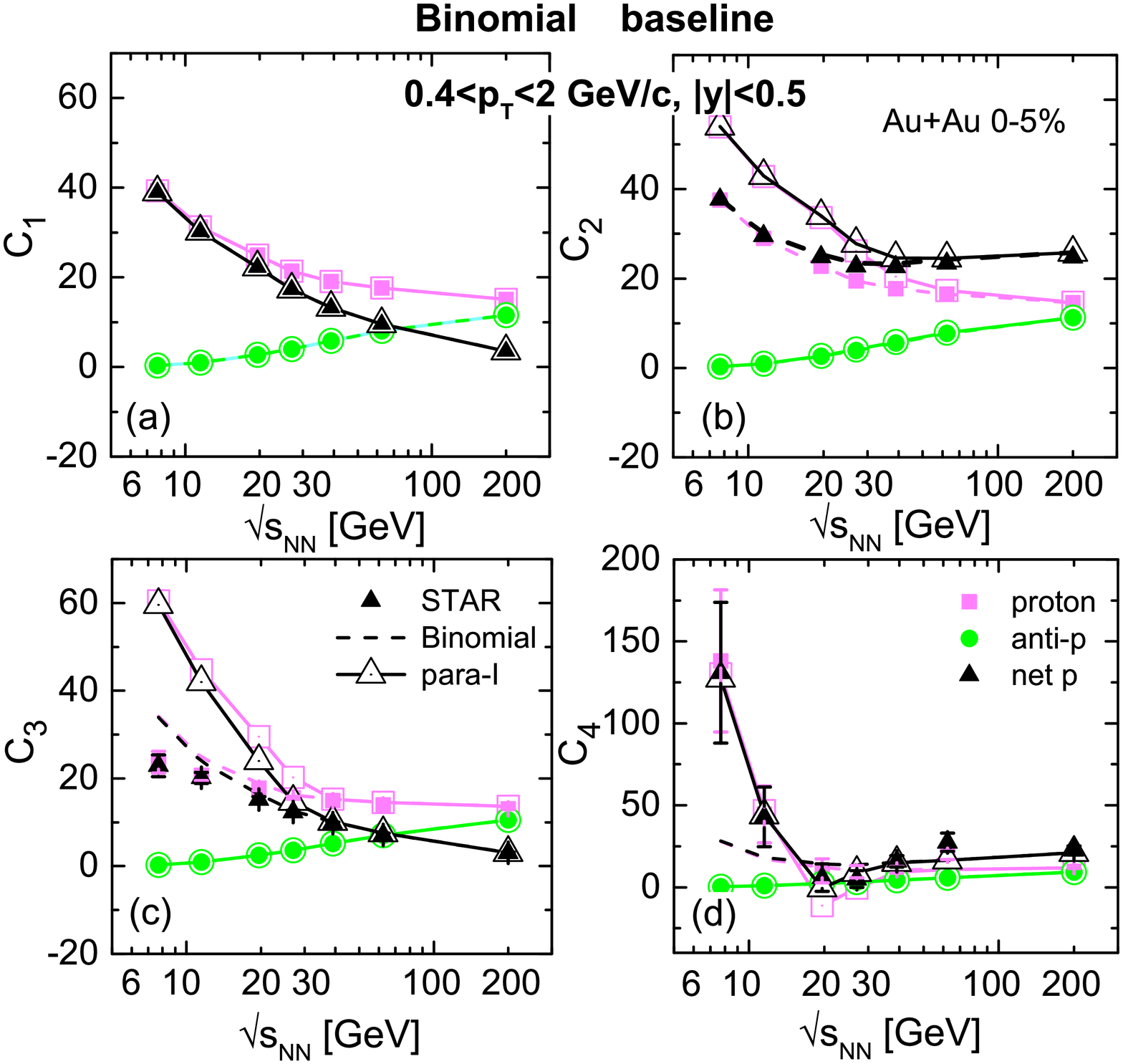} 
\caption{(Color online) Energy dependence of cumulants $C_1\sim C_4$ for protons, anti-protons and
net protons, with Poisson baselines (left panels) and Binomial baselines (right panels).}
\label{cumulant4-0005-appendix}
\end{figure*}

\vspace{-0.10in}
\textbf{\quad A. critical fluctuations in static and infinite system}.
\\

The standard critical fluctuations in a static and infinite medium were firstly
deduced in~\cite{Stephanov:1999zu,Stephanov:2008qz} through considering
a joint probability distribution for the
sigma field and the occupation numbers $n_p$ of the produced particles. In this paper, we assume the coupling between the sigma field and the classical
particles introduces a variable effective mass for the particle, which gives a modified distribution function
that strongly fluctuates near the critical point.  In
appendix A, we will show that, for a static and infinite medium,
we could re-derive the early results of ~\cite{Stephanov:2008qz}
with such modified distribution function.

For a static and infinite medium, integrating Eqs.(3-5) over the whole position space gives:
\begin{equation}
\left\langle \delta n_{p_1}...\delta n_{p_n}\right\rangle =\prod_{i=1...n}\left(
\int d^{3}x_{i}\right) ^{n}\left\langle \delta f_{1}...\delta
f_{n}\right\rangle _{c}/\prod_{i=1...n}\left( \int d^{3}x_{i}1\right) ^{n}.
\end{equation}%

The two point correlator $\left\langle \delta n_{p_1}\delta n_{p_2}\right\rangle_c$
involves an integration of the equal-time propagator of the sigma field:
\begin{equation}
\int d^{3}x_{1}d^{3}x_{2} D\left( x_{1}-x_{2}\right)=\int
d^{3}R\int d^{3}r\frac{1}{4\pi r}e^{-m_\sigma r}=\frac{V}{m_\sigma^{2}}
\end{equation}%
with $R=\frac{x_{1}+x_{2}}{2},r=x_{2}-x_{1}$ and $D\left( x_{1}-x_{2}\right) =\frac{1}{4\pi r}e^{-m_\sigma r}$, thus
\begin{equation}
\left\langle \delta n_{p_1}\delta n_{p_2}\right\rangle_c=\frac{f_{01}f_{02}}{%
\omega _{p_1}\omega _{p_2}}\frac{G^{2}}{T}\frac{V}{m_\sigma^{2}}.
\end{equation}%
where $G=gm_p$ and $\omega_{p}=p_\mu u^\mu$ for a static system.

The 3-point correlator of the sigma field can also be calculated with the variables
substitution: $x_{1}-z=r_{1},\;x_{2}-z=r_{2},\;x_{3}-z=r_{3},\;\frac{x_{1}+z%
}{2}=R$, and the Jacobi determination $\frac{\partial \left(
r_{1,}r_{2},r_{3},R\right) }{\partial \left( x_{1},x_{2},x_{3},z\right) }=1$.
The integration of the $3$-point correlator can be written as:
\begin{eqnarray}
&&\int d^{3}x_{1}d^{3}x_{2}d^{3}x_{3}\int d^{3}zD\left( x_{1}-z\right)
D\left( x_{2}-z\right) D\left( x_{3}-z\right)  \notag \\
&=&\int d^{3}r_{1}d^{3}r_{2}d^{3}r_{3}\int d^{3}RD\left( r_{1}\right)
D\left( r_{2}\right) D\left( r_{3}\right)  \notag \\
&=&V\left( \frac{1}{m_\sigma^{2}}\right) ^{3}.
\end{eqnarray}%
Then, the 3-point correlator of particle density in momentum space is
\begin{equation}
\left\langle \delta n_{p_1}\delta n_{p_2}\delta n_{p_3}\right\rangle _{c}=\frac{%
2\lambda _{3}}{V^{2}T}\frac{f_{01}f_{02}f_{03}}{\omega _{p_1}\omega _{p_2}\omega
_{p_3}}\left( \frac{G}{m_\sigma^{2}}\right) ^{3}.
\end{equation}%
Following the same steps, we can calculate the 4-particle
correlator in momentum space with the variables substitution, which gives
\begin{eqnarray}
&&\left\langle \delta n_{p_{1}}\delta n_{p_{2}}\delta n_{p_{3}}\delta
n_{p_{4}}\right\rangle _{c}  \notag \\
&=&\frac{6}{V^{3}T}\frac{f_{01}f_{02}f_{03}f_{04}}{%
\omega _{p_1}\omega _{p_2}\omega _{p_3}\omega _{p_4}}\left( \frac{G}{%
m_\sigma^{2}}\right) ^{4}\left[ 2\left( \frac{\lambda _{3}}{m_\sigma}\right) ^{2}-\lambda
_{4}\right] .
\end{eqnarray}
With Eqs.(18-19), we now reproduce the static fluctuations for a static and infinite medium
once deduced in paper~\cite{Stephanov:2008qz}.\\[0.10in]

\textbf{\quad B. cumulants of protons anti-protons and net protons}.
\\

In Fig.~1-4, we only present the cumulants of net protons for clearness of the figures.
In appendix B, we will compare the cumulants of protons, anti-protons and net protons
in central Au+Au collisions within $0.4<p_T<2 \ \mathrm{GeV}$.  Fig.~7
shows the energy dependence of these cumulants with either Poisson or Binomial
baselines. For clearness, we only choose one parameter set (para-I)
to calculate the critical fluctuations. For anti-protons, the cumulants with both critical and non-critical
 fluctuations (para-I) are almost overlap with non-critical fluctuation baselines due to the small mean values
at various collision energies which largely suppress the critical fluctuations.
Basically, the cumulants of net protons follows the trends of protons. As already discovered in Fig.~1
and Fig.~2, our calculations can roughly fit $C_{4}$  for protons and anti-protons with pare-I,
but over-predict $C_{2}$ and $C_{3}$ due to the positive contributions from the critical
fluctuations.\\[0.10in]

\textbf{\quad C. momentum dependence of critical fluctuations}.
\\

\begin{figure*}[tbp]
\centering
\includegraphics[width=5.6 in]{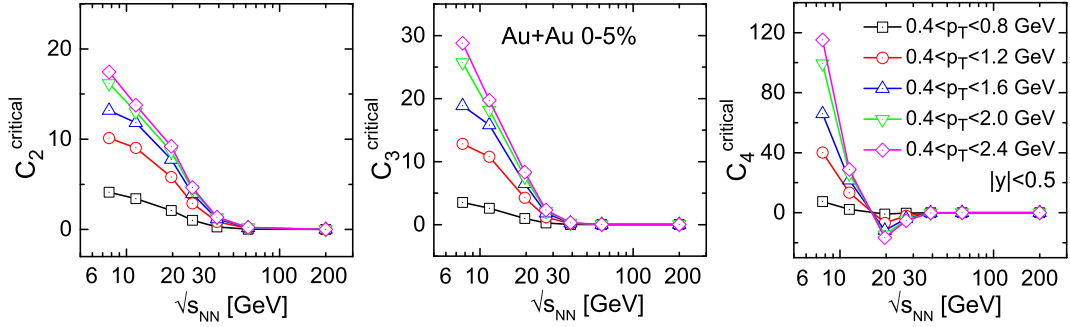}
\caption{(color online) $p_T$ acceptance dependence of $C_2^{critical}, \ C_3^{critical}$ and $C_4^{critical} $ of net protons.}
\label{Fig8}
\end{figure*}



In this appendix C, we explore the acceptance dependence of critical fluctuations in the transverse momentum space.
The critical fluctuations in our calculations are mainly influenced by two factors:
the mean value (average number) of net protons within specific acceptance window and the correlation length $\xi$.
Fig.~8 shows the collision energy dependent cumulants $C_2^{critical}, \ C_3^{critical}$ and $\ C_4^{critical} $ of net protons,
calculated from Eqs.(11-13) within different $p_T$ windows and within momentum rapidity $|y|<0.5$.
At larger collision energies above 39 GeV, all cumulants almost approach zero
because of the smaller correlation length $\xi$ and reduced mean values of net protons within specific acceptance window.
At lower collision energies, $C_n^{critical} \ (n=2,3,4)$ increases as the $p_T$ window is broadened.
In fact, the mean value of net protons increases with the increase of the $p_T$-acceptance,
leading to the enhanced signals of the critical fluctuations. Note that with the acceptance increased
to $0.4<p_T<2 \ \mathrm{GeV}$, $C_n^{critical} \ (n=2,3,4)$ almost reach saturations
since most of protons and anti-protons are produced below 2 GeV.
Fig.~8 also shows that $C_n^{critical} \ (n=2,3,4)$ gradually increase as the collision energy decreases.
Meanwhile, the correlation length $\xi$ in table I decrease from O(3fm) to 1 fm.
As discussed in Sec. V, both correlation length and the
mean value of net protons $C_1$ influence the critical fluctuations. At lower collision energy,
the drastically increased $C_1$ becomes the dominant factor, which leads to the increasing trend of
$C_n^{critical} \ (n=2,3,4)$ with the decrease of the collision energy.

In a recent paper~\cite{Ling:2015yau}, Ling and Stephanov have investigated the
acceptance dependence of critical fluctuations in both transverse momentum and rapidity windows,
using a simplified freeze-out surface constructed from the blast-wave model.
They demonstrated that  the $p_T$ window dependence is significant
for different order cumulants, which is qualitatively agree with what we find in Fig.~8.
They also found that, for typical experimental rapidity acceptance window
$\Delta y \leq 1$, larger acceptance leads to significantly increased critical fluctuations.
In this appendix, we will not further explore such rapidity acceptance dependence
since the 3+1-d freeze-out surfaces used in our calculations
are constructed from the 2+1-d freeze-out surfaces of {\tt VISH2+1} using the
momentum rapidity and space rapidity correlations.


\noindent\textbf{Acknowledgments}
\\[-0.20in]

We thank M. Asakawa, A. Dumitru, D. Hou, S. Gupta, Y. Liu, X. Luo, M. Stephanov,  N. Xu
for valuable discussions. This work is supported by the NSFC and the MOST under grant
Nos. 11435001 and 2015CB856900.

\end{document}